\documentclass[10pt]{amsart}
\usepackage{amssymb}
\usepackage{enumerate}
\usepackage{epsf}
\usepackage{psfrag}
\DeclareGraphicsExtensions{.eps,.art,.ART,.ps}
\newcommand{\bbR}{\mathbb{R}}      % real numbers %
\newcommand{\grad}{\operatorname{grad}}

\newcommand{\dive}{\operatorname{div}}
\newcommand{\diag}{\operatorname{diag}}
\begin{document}
\begin{abstract}
We show that the equation of motion for a rigid one-dimensional elastic body (i.e.~a rod or string whose speed of sound is equal to the speed of light) in a two-dimensional spacetime is simply the wave equation. We then solve this equation in a few simple examples: a rigid rod colliding with an unmovable wall, a rigid rod being pushed by a constant force, a rigid string whose endpoints are simultaneously set in motion (seen as a special case of Bell's spaceships paradox), and a radial rigid string that has partially crossed the event horizon of a Schwarzschild black hole while still being held from the outside.
\end{abstract}
%
%%%%%%%%%%%%%%%%%%%%%%%%%%%% Definition of title page %%%%%%%%%%%%%%%%%%%%%%%%%%%-
%
\title{Relativistic elasticity of rigid rods and strings}
\author{Jos\'{e} Nat\'{a}rio}
\address{Center for Mathematical Analysis, Geometry and Dynamical Systems, Mathematics Department, Instituto Superior T\'ecnico, Universidade de Lisboa, Portugal}
%\subjclass[2000]{Primary 53078}
%\thanks{Partially funded by FCT/Portugal through project PEst-OE/EEI/LA0009/2013.}
%
%
\maketitle
%
%
%%%%%%%%%%%%%%%%%%%%%%%%%%%%%%%%%%% Section 0 %%%%%%%%%%%%%%%%%%%%%%%%%%%%%%%%%%
%
\section*{Introduction}
Relativistic elasticity in its modern form was originally formulated by Carter and Quintana \cite{CQ72}, and further developed in \cite{Maugin78, KM92, Tahvildar98, BS02, KS03} (see also \cite{Wernig06}). It has been used to compute the speed of sound in relativistic solids \cite{Carter73, Bento85} and to study elastic equilibrium states \cite{Park00, KS04, ABS08, ABS09, BCV10, BCMV12, AC14} or dynamical situations \cite{Magli97, Magli98, CH07}.

Here we consider one-dimensional elastic bodies moving in two-dimensional spacetimes (typically totally geodesic Lorentzian submanifolds of four-dimensional spacetimes). In this case the relativistic elasticity theory is much simpler than in the general case, and we develop it {\em ab initio} for the convenience of the reader. If we further require our elastic bodies to be {\em rigid}, that is, to have a speed of sound equal to the speed of light, then we are led to a unique elastic model (considered previously, albeit in a non-geometric framework, in \cite{HM52, McCrea52, BF03}). Moreover, the equation of motion turns out to be simply the wave equation, which in two dimensions has an explicit solution when written in conformal coordinates. This allows us to obtain closed form solutions for many interesting problems.

The organization of the paper is as follows: in Section~\ref{section1} we develop the general theory of one-dimensional elastic bodies moving in two-dimensional spacetimes, and find the equation of motion for {\em rigid} elastic bodies. There is an interesting connection with conformal coordinate systems, which we exploit to obtain static solutions in general static spacetimes and expanding solutions in general cosmological spacetimes. In Section~\ref{section2} we solve the wave equation for the problem of a semi-infinite rigid rod in Minkowski spacetime colliding with an unmovable wall. This basic result is then used in Section~\ref{section3} to construct the solution representing a {\em finite} rigid rod colliding with an unmovable wall. We note that this can be used to analyze the celebrated car and garage paradox, since we can now model what happens to the car after colliding with the garage's back wall. The same system, conveniently reinterpreted, is used in Section~\ref{section4} to describe a rigid rod being pushed by a constant force. In Section~\ref{section5} we piece together different versions of the basic solution to represent a rigid string whose endpoints are simultaneously set in motion. This can be seen as a special case of the celebrated Bell spaceships paradox, where now we can model exactly what happens to the string connecting the two spaceships. Finally, in Section~\ref{section6} we use the conformal Kruskal-Szekeres coordinates to study a radial rigid string that has partially crossed the event horizon of a Schwarzschild black hole while still being held from the outside.

We follow the conventions of \cite{MTW73}; in particular, we use a system of units for which $c=G=1$.
%
%
%%%%%%%%%%%%%%%%%%%%%%%%%%%%%%%%%%% Section 1 %%%%%%%%%%%%%%%%%%%%%%%%%%%%%%%%%%
%
\section{String theory}\label{section1}
We will be concerned with one-dimensional elastic bodies moving in a two-dimensional spacetime $(M,g)$ (typically a totally geodesic Lorentzian submanifold of a four-dimensional spacetime). Such objects may be thought of as either rods or strings (depending on the situation), but for the sake of definiteness we will call them strings in this section.

In the spacetime (or {\em Eulerian}) picture, a string can be described by a scalar field $\lambda:M \to \bbR$ with nonvanishing gradient, whose level sets are the worldlines of the string particles\footnote{Note the difference between these macroscopic strings and the (pre-quantized) strings of string theory, whose action is invariant under reparameterizations: for the latter one does not keep track of the motion of their individual points, and so they have no degrees of freedom in a $2$-dimensional spacetime.}. The local rest spaces of these particles are determined by the gradient of $\lambda$, and so their two-velocity is
\begin{equation} \label{u}
u^\alpha = \frac1n \varepsilon^{\alpha \beta} \nabla_\beta \lambda,
\end{equation}
where $\varepsilon_{\alpha \beta}$ is the two-dimensional volume element (with an orientation choice such that $u^\alpha$ is future-pointing) and $n^2 = \left|\grad \lambda\right|^2 = \nabla_\alpha \lambda \, \nabla^\alpha \lambda$. We assume that $n^2 = 1$ for the unstretched string, so that $s=1/n$ is the local stretch factor. Since we are in two dimensions, the (non-null) stress-energy tensor of the string is necessarily of the perfect fluid form
\begin{equation}
T^{\alpha\beta} = (\rho + p) u^\alpha u^\beta + p g^{\alpha \beta},
\end{equation}
and we assume that the energy density $\rho=\rho(n)$ and the pressure\footnote{Since we are dealing with a two-dimensional spacetime, this pressure is actually a force -- tension ($p<0$) or compression ($p>0$) of the string.} $p=p(n)$ are functions of $n$ only. From energy-momentum conservation we obtain
\begin{equation}
\nabla_\alpha T^{\alpha\beta} = 0 \Leftrightarrow \left[ \nabla_u (\rho + p) + (\rho + p) \dive u \right] u^\beta + (\rho + p) \nabla_uu^\beta + \nabla^\beta p = 0,
\end{equation}
which upon contracting with $u_\beta$ yields
\begin{equation} \label{p0}
p = - \frac{\nabla_u \rho}{\dive u} - \rho = - \left( \frac{\nabla_u n}{\dive u} \right) \left( \frac{d \rho}{dn} \right) - \rho.
\end{equation}
Taking the divergence of \eqref{u} leads to
\begin{equation}
\dive u = - \frac{\nabla_\alpha n}{n^2} \varepsilon^{\alpha \beta} \nabla_\beta \lambda + \frac1n \varepsilon^{\alpha \beta} \nabla_\alpha \nabla_\beta \lambda = - \frac1n \nabla_u n
\end{equation}
(since the volume element is covariantly constant and $\nabla_\alpha \nabla_\beta \lambda$ is symmetric), and so \eqref{p0} becomes
\begin{equation} \label{p}
p = n \frac{d \rho}{dn} - \rho.
\end{equation}
Therefore the choice of the function $\rho=\rho(n)$ (which we will call the {\em elastic law}) completely fixes the pressure $p=p(n)$. Although there are many possible choices for elastic laws, a particularly simple one is that leading to a {\em rigid} string, that is, a string for which the speed of sound is equal to the speed of light. This condition can be written as \cite{Christodoulou97}
\begin{equation}
\frac{d p}{d \rho} = 1 \Leftrightarrow \rho = p + \rho_0,
\end{equation}
where $\rho_0 > 0$ is a constant (the string's linear rest density when relaxed). Using \eqref{p} we obtain
\begin{equation}
n \frac{d \rho}{dn} = 2\rho - \rho_0,
\end{equation}
which is easily integrated to give
\begin{equation} \label{rhorigid}
\rho = \frac{\rho_0}2 \left(n^2 + 1 \right)
\end{equation}
(assuming $p = 0$, hence $\rho=\rho_0$, for $n=1$). Equation~\eqref{p} then yields
\begin{equation} \label{prigid}
p = \frac{\rho_0}2 \left(n^2 - 1 \right).
\end{equation}
Note that $\rho \geq |p|$, that is, the dominant energy condition is satisfied. Equations~\eqref{rhorigid} and \eqref{prigid}, first obtained by Hogarth and McCrea \cite{HM52, McCrea52}, completely characterize the rigid string. To obtain its equation of motion we write the energy-momentum tensor as
\begin{equation}
T_{\alpha\beta} = \frac{\rho_0}2 \left[\left( n^2 + 1 \right) u_\alpha u_\beta + \left( n^2 - 1 \right) \frac1{n^2} \nabla_\alpha \lambda \, \nabla_\beta \lambda \right].
\end{equation}
Using the identity\footnote{This identity is immediate from the fact that $u$ and $\frac1n \grad \lambda$ form an orthonormal basis; it can also be obtained by direct substitution using $\varepsilon_{\alpha \beta}\varepsilon_{\mu \nu} = - g_{\alpha \mu} g_{\beta \nu} + g_{\alpha \nu} g_{\beta \mu}$ and $n^2 = \nabla_\alpha \lambda \, \nabla^\alpha \lambda$.}
\begin{equation}
g_{\alpha \beta} = - u_\alpha u_\beta + \frac1{n^2} \nabla_\alpha \lambda \, \nabla_\beta \lambda,
\end{equation}
we find
\begin{equation}
T_{\alpha\beta} = \rho_0 \left( \nabla_\alpha \lambda \, \nabla_\beta \lambda - \frac12 \nabla_\mu \lambda \, \nabla^\mu \lambda \, g_{\alpha \beta} - \frac12 g_{\alpha \beta} \right),
\end{equation}
which, apart from the covariantly constant term proportional to $g_{\alpha \beta}$, is just the energy-momentum tensor for the massless scalar field. As is well known (and easily checked), we have
\begin{equation}
\nabla^\alpha T_{\alpha\beta} = 0 \Leftrightarrow  \nabla_\beta \lambda \, \Box \lambda = 0,
\end{equation}
and so, since $\grad \lambda \neq 0$, we see that the equation of motion for the rigid string is simply the wave equation\footnote{It may seem that the requirement that $\grad \lambda$ be spacelike will render this equation effectively nonlinear; however, it can be checked from the general solution~\eqref{wave_soln} of the wave equation in one spatial dimension (valid for general harmonic coordinates) that the zeros of $|\grad \lambda|^2$ occur along null lines, and so if $\grad \lambda$ is spacelike on the initial data then it is automatically spacelike on the corresponding domain of dependence.}
\begin{equation} \label{wave_eqn}
\Box \lambda = 0.
\end{equation}

Since $\grad \lambda$ is nonvanishing and spacelike, $\lambda$ forms part of a conformal system of (necessarily harmonic) coordinates. Indeed, it is possible (assuming $M$ simply connected) to find a function $\tau:M \to \bbR$ such that
\begin{equation}
\nabla_\alpha \tau = \varepsilon_{\alpha \beta} \nabla^\beta \lambda,
\end{equation}
because the one-form $\varepsilon_{\alpha \beta} \nabla^\beta \lambda$ is closed:
\begin{equation}
\varepsilon^{\alpha \beta} \nabla_\alpha \left( \varepsilon_{\beta \mu} \nabla^\mu \lambda\right) = \varepsilon^{\alpha \beta}  \varepsilon_{\beta \mu} \nabla_\alpha \nabla^\mu \lambda = \delta^{\alpha}_{\,\,\,\, \mu} \nabla_\alpha \nabla^\mu \lambda = \Box \lambda = 0,
\end{equation}
where we used
\begin{equation}
\varepsilon^{\alpha \beta}  \varepsilon_{\beta \mu} = \delta^{\alpha}_{\,\,\,\, \mu}.
\end{equation}
Moreover, we have
\begin{equation}
\Box \tau = \nabla^\alpha \nabla_\alpha \tau = \nabla^\alpha \varepsilon_{\alpha \beta} \nabla^\beta \lambda = \varepsilon_{\alpha \beta} \nabla^\alpha \nabla^\beta \lambda = 0,
\end{equation}
and so the function $\tau$ also satisfies the wave equation. Finally,
\begin{equation}
\nabla_\alpha \tau \, \nabla^\alpha \tau = \varepsilon_{\alpha \beta} \varepsilon^{\alpha \mu} \nabla^\beta \lambda \, \nabla_\mu \lambda = - \delta_\beta^{\,\,\,\, \mu} \nabla^\beta \lambda \, \nabla_\mu \lambda = - n^2,
\end{equation}
and so the metric $g$ in the harmonic coordinates $(\tau, \lambda)$ is written
\begin{equation} \label{harmonic}
ds^2 = \frac1{n^2} \left( - d\tau^2 + d\lambda^2 \right).
\end{equation}
This provides an interesting interpretation for conformal coordinates in two-dimensional spacetimes: they can be thought of as being associated to a rigid string, with the time coordinate determining its simultaneity lines, the spatial coordinate determining the worldlines of its points, and the conformal factor yielding the square of the stretch factor. 

For static spacetimes, for instance, we can always find a coordinate system such that the metric is written in the form
\begin{equation}
ds^2 = - e^{2\phi(x)} dt^2 + dx^2.
\end{equation}
Defining the new spatial coordinate
\begin{equation}
y = \int e^{-\phi(x)} dx
\end{equation}
we then have
\begin{equation}
ds^2 = e^{2\phi(y)} (- dt^2 + dy^2 ),
\end{equation}
and so there exist static rigid strings with stretch factor given by $e^\phi$. This can be readily used to construct static solutions of the wave equation \eqref{wave_eqn} describing, for instance, radially hanging rigid strings in the Schwarzschild spacetime.

As a second example, cosmological solutions often admit totally geodesic $2$-dimensional Lorentzian submanifolds with line element
\begin{equation}
ds^2 = - dt^2 + a^2(t) dx^2.
\end{equation}
Introducing the conformal time coordinate
\begin{equation}
\tau = \int \frac{dt}{a(t)}
\end{equation}
leads to
\begin{equation}
ds^2 = a^2(\tau) ( - d \tau^2 + dx^2 ),
\end{equation}
and so there exist rigid strings whose points follow the Hubble flow and whose stretch factor is, as one would expect, the cosmological radius $a$.

As an aside, we note that it is possible to extend part of the above results to strings with arbitrary (constant) speed of sound $c < 1$: the density and pressure are in this case given by\footnote{The dominant energy condition still holds in this case, as is easily seen if one writes it in the form $0 \leq n\frac{d\rho}{dn} \leq 2 \rho$.}
\begin{align*}
& \rho = \frac{\rho_0}{1+c^2}\left( n^{c^2 + 1} + c^2 \right), \\
& p =  \frac{\rho_0 c^2}{1+c^2}\left( n^{c^2 + 1} - 1 \right),
\end{align*}
and $\lambda$ satisfies the nonlinear wave equation
\begin{equation} \label{wave_nonlinear}
\Box \lambda + \frac{c^2 - 1}{n^2} \nabla^\alpha \lambda \, \nabla^\beta \lambda \, \nabla_\alpha \nabla_\beta \lambda = 0.
\end{equation}
This equation can be thought of as a wave equation for an {\em acoustic metric} $h$, whose inverse $h^{-1}$ is determined by
\begin{equation}
h^{\alpha\beta} = g^{\alpha\beta} + \frac{c^2 - 1}{n^2} \nabla^\alpha \lambda \, \nabla^\beta \lambda.
\end{equation}
Note that in the string's local orthonormal rest frames (with respect to the spacetime metric $g$) the inverse acoustic metric $h^{-1}$ is represented by $\diag(-1,c^2)$.
%
%
%%%%%%%%%%%%%%%%%%%%%%%%%%%%%%%%%%% Section 2 %%%%%%%%%%%%%%%%%%%%%%%%%%%%%%%%%%
%
\section{Hitting a wall}\label{section2}
It is instructive to analyze simple motions of rigid rods or strings in familiar spacetimes. As a first example, we consider a semi-infinite rod in Minkowski spacetime, initially relaxed and moving with velocity $v$ along the $x$-axis, before colliding with an unmovable wall located at $x=0$. If we set
\begin{equation}
d \lambda = \dot{\lambda} \, dt + \lambda ' dx
\end{equation}
then velocity of the points in the rod is given by
\begin{equation}
d \lambda = 0 \Leftrightarrow \frac{dx}{dt} = - \frac{\dot{\lambda}}{\lambda '},
\end{equation}
and so before the collision we must have
\begin{equation}
- \frac{\dot{\lambda}}{\lambda '} = v  \qquad \text{ and } \qquad - \dot{\lambda}^2 + (\lambda ')^2 = 1,
\end{equation}
that is,
\begin{equation}
\dot{\lambda} = - \gamma v \qquad \text{ and } \qquad \lambda ' = \gamma,
\end{equation}
where 
\begin{equation}
\gamma=\frac1{\sqrt{1-v^2}}
\end{equation}
is the usual Lorentz factor. Assuming that the collision occurs at $t=0$, we must then solve the following initial-boundary value problem:
\begin{equation} \label{problem_wall}
\begin{cases}
\Box \lambda = 0 \quad & (t>0,x<0)\\
\lambda(0,x) = \gamma x \quad & (x<0)\\
\dot{\lambda}(0,x) = - v\gamma \quad & (x<0)\\
\lambda(t,0) = 0 \quad & (t>0)
\end{cases}
\end{equation}

\begin{figure}[h!]
\begin{center}
\psfrag{t}{$t$}
\psfrag{x}{$x$}
\psfrag{I}{$I$}
\psfrag{II}{$II$}
\epsfxsize=.5\textwidth
\leavevmode
\epsfbox{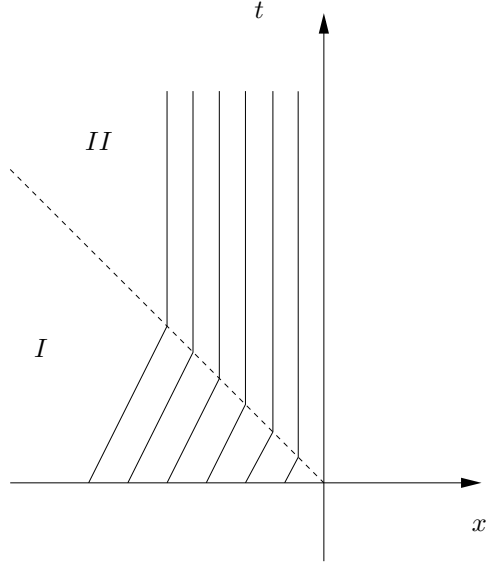}
\end{center}
\caption{Spacetime diagram for the collision of a semi-infinite rigid rod with an unmovable wall.}\label{wall}
\end{figure}

Using the general form of the solution of the wave equation in one spatial dimension\footnote{Note that this may lead to weak (not even $C^1$) solutions of the wave equation; such solutions are not physically problematic, since energy-momentum conservation holds.},
\begin{equation} \label{wave_soln}
\lambda(t,x) = f(x-t) + g(x+t),
\end{equation}
one easily obtains from the initial condition
\begin{equation}
f(x) = \frac12 \gamma (1+v) x \qquad \text{ and } \qquad g(x) = \frac12 \gamma (1-v) x \qquad (x < 0),
\end{equation}
and from the boundary condition
\begin{equation}
g(t) = - f(-t) =  \frac12 \gamma (1+v) t  \qquad (t > 0).
\end{equation}
Therefore the solution to \eqref{problem_wall} is
\begin{equation}
\lambda(t,x) = 
\begin{cases}
\gamma (x - vt) \qquad & (t>0,x<0,x + t < 0) \\
\gamma (1+v) x  \qquad & (t>0,x<0,x + t > 0)
\end{cases}
\end{equation}
The worldlines of the particles in the rod (level sets of $\lambda$) are depicted in Figure~\ref{wall}. We see that there is a compression wave propagating from the collision event $(t,x)=(0,0)$ through the rod at the speed of light, represented by the line $x+t=0$. Before being reached by the compression wave (region $I$ in Figure~\ref{wall}), the points in the rod are moving with velocity $v$ and are not compressed:
\begin{equation}
n^2 = | d\lambda |^2 = - v^2 \gamma^2 + \gamma^2 = 1.
\end{equation}

\noindent After being reached by the compression wave (region $II$ in Figure~\ref{wall}), the points of the rod are at rest with respect to the wall and compressed:
\begin{equation}
n^2 = | d\lambda |^2 = \gamma^2 (1 + v)^2 = \frac{(1 + v)^2}{1 - v^2} = \frac{1 + v}{1 - v} > 1,
\end{equation}
and so
\begin{equation}
p = \frac{\rho_0}2 (n^2 - 1) = \frac{\rho_0 v}{1 - v} > 0.
\end{equation}
Note that the difference between the unstressed region $I$ and the constantly stressed region $II$ is merely the spacing between the (straight) level sets of $\lambda$. From \eqref{harmonic} it is clear that constant stress is only possible in flat spacetime (where there are no tidal forces), and that in this case $\lambda$ is necessarily an affine function of $(t,x)$.

It is possible to construct similar solutions for \eqref{wave_nonlinear}, corresponding to rods with constant speed of sound $c<1$. In these solutions, the compression wave travels with speed
\begin{equation}
w = c + \frac12 (c^2 - 1) v + \frac1{24c} (c^2 - 1) (5c^2 - 3) v^2 + \ldots
\end{equation}
in the rest frame of the wall.
%
%
%%%%%%%%%%%%%%%%%%%%%%%%%%%%%%%%%%% Section 3 %%%%%%%%%%%%%%%%%%%%%%%%%%%%%%%%%%
%
\section{Car and garage paradox}\label{section3}
\begin{figure}[h!]
\begin{center}
\psfrag{t}{$t$}
\psfrag{x}{$x$}
\epsfxsize=.5\textwidth
\leavevmode
\epsfbox{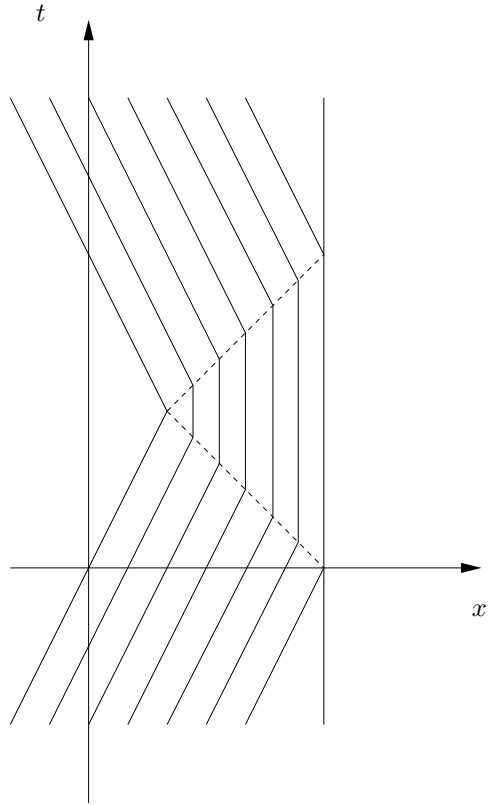}
\end{center}
\caption{Spacetime diagram for the collision of a finite rigid rod with an unmovable wall.}\label{car}
\end{figure}

The solution of the previous section can easily be adapted to model the collision of a finite rod with an unmovable wall. Such collisions were analyzed, albeit in a non-geometric framework, in \cite{HM52, McCrea52, BF03}. The corresponding spacetime diagram is depicted in Figure~\ref{car}, where we assume that the wall is placed at $x=l$, the rod has proper length $l_0=\gamma l$, and the collision occurs at $t=0$. By symmetry, it is clear that as the compression wave reaches the free end of the rod a rarefaction wave starts propagating backwards\footnote{Note that this allows us to sidestep the highly nonlinear condition $n^2 \equiv -\dot{\lambda}^2 + {\lambda'}^2 = 1$ at the free boundary.}, and the points of the rod start moving away from the wall with velocity $-v$. The rod's length at the instant $t=\frac{l}{1+v}$, when it is fully compressed, is easily computed to be
\begin{equation} \label{l_c}
l_c = \frac{l}{1+v} = l_0 \sqrt{\frac{1-v}{1+v}} \equiv \frac{l_0}{n}.
\end{equation}

This offers an elastic solution to the celebrated car and garage paradox, where a car longer than a garage is moving fast enough so that it fits inside (by length contraction in the garage's frame). The (apparent) paradox is, of course, that in the car's rest frame it is the garage which is contracted, and so the car cannot possibly fit inside. As is well known, the resolution of this paradox hinges on the different notions of simultaneity: in the car's frame, the event where the front of the car hits the garage's back wall occurs {\em before} the event where the rear of the car passes through the door. The car cannot stop instantaneously at the moment when it hits the wall; if it is rigid then a compression wave will stop it momentarily, as shown in Figure~\ref{car}. Assuming the garage's door to be located at $x=0$, so that the contracted length $\frac{l_0}{\gamma}$ of the car is precisely equal to the size $l$ of the garage, then the car is completely inside (and relaxed) at time $t=0$, as seen in the garage's frame. In the car's (initial) frame, however, the set of events simultaneous with the rear of the car going through the door is the line $t = v x$; thus, in this frame, the front of the car is already compressed when the rear enters the garage.

Similar solutions for finite rods with constant speed of sound $c<1$ can be constructed from the corresponding collisions of semi-infinite rods with an unmovable wall; solutions of this kind were considered in \cite{WS07}.
%
%
%%%%%%%%%%%%%%%%%%%%%%%%%%%%%%%%%%% Section 4 %%%%%%%%%%%%%%%%%%%%%%%%%%%%%%%%%%
%
\section{Pushing a rigid rod}\label{section4}
\begin{figure}[h!]
\begin{center}
\psfrag{t}{$t$}
\psfrag{x}{$x$}
\epsfxsize=.5\textwidth
\leavevmode
\epsfbox{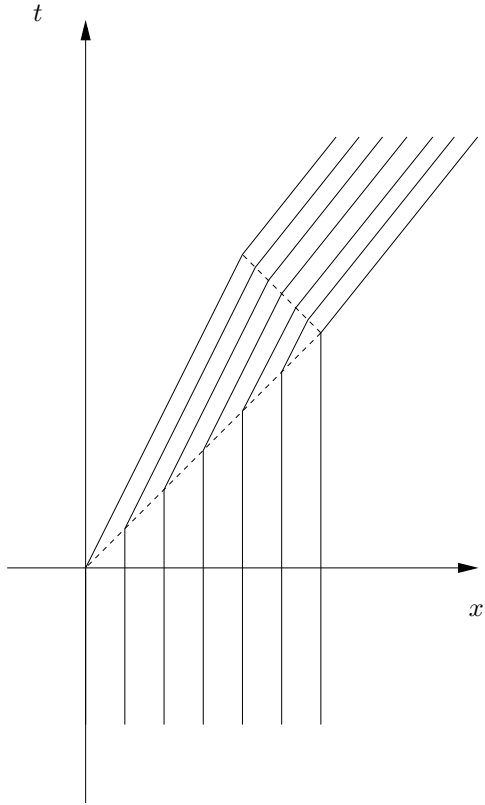}
\end{center}
\caption{Spacetime diagram for a finite rod being pushed by a constant force for a finite amount of time.}\label{rod}
\end{figure}

By changing to the rod's initial rest frame (and reversing the $x$-axis), we can reinterpret the solution above as describing a finite rod being pushed by a constant force for a finite amount of time and then released. Consider, as illustrated in Figure~\ref{rod}, a finite rod in Minkowski spacetime, initially relaxed and at rest. At time $t=0$ a constant force 
\begin{equation}
p = \frac{\rho_0 v}{1 - v}
\end{equation}

\begin{figure}[h!]
\begin{center}
\psfrag{t}{$t$}
\psfrag{ts}{$t_s$}
\psfrag{l}{$l$}
\psfrag{l0}{$l_0$}
\psfrag{lc}{$l_c$}
\epsfxsize=.6\textwidth
\leavevmode
\epsfbox{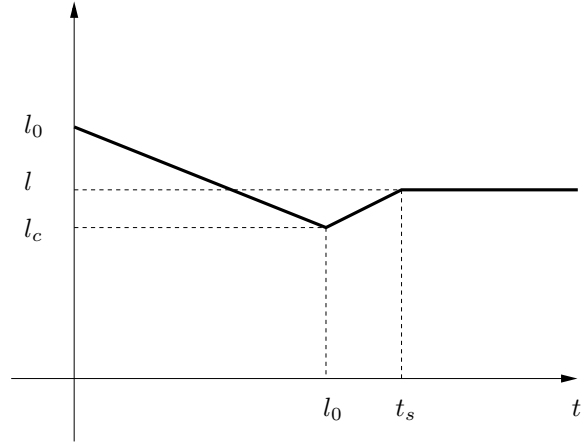}
\end{center}
\caption{Graph for the rod's length as a function of time.}\label{graph}
\end{figure}

\noindent starts to be exerted at the rear end, so that this end starts moving with velocity $v$. At the same instant, a compression wave propagates along the rod at the speed of light, setting all points it reaches in motion with the same velocity $v$. As the compression wave reaches the front end of the rod, a rarefaction wave starts propagating backwards, also at the speed of light, setting the points it reaches to move with velocity
\begin{equation}
w = \frac{2v}{1 + v^2},
\end{equation}
while returning the rod to its relaxed state. If the force ceases as the rarefaction wave reaches the rear end, that is, at time
\begin{equation}
t_s = \frac{2l_0}{1 + v},
\end{equation}
then the rod is left relaxed and moving with velocity $w$, so that its length is now
\begin{equation}
l = l_0 \sqrt{1 - w^2} = \frac{1-v^2}{1+v^2} \, l_0.
\end{equation}
It is interesting to note that the rod's length at the instant $t=l_0$ when completely compressed is
\begin{equation}
l_c = l_0 (1 - v) \equiv \frac{l_0}{n \gamma} < l.
\end{equation}
The rod's length (as measured in its initial rest frame) is plotted in Figure~\ref{graph} as a function of time. This offers a picture of how Lorentz contraction is attained in a physically realistic setting.

If the force keeps acting on the rear end then the rod will repeat the same cycle of compression and rarefaction in its new rest frame. After $N$ cycles it will be moving with speed
\begin{equation}
w = \tanh(2Nu),
\end{equation}
where we set
\begin{equation}
v = \tanh(u).
\end{equation}
The duration of each cycle, as measured by an observer comoving with the rear end, can be seen from \eqref{l_c} to be
\begin{equation}
2l_0 \sqrt{\frac{1-v}{1+v}} = 2l_0 e^{-u},
\end{equation}
and so the proper time measured by this observer after $k$ cycles is
\begin{equation}
\tau = 2Nl_0 e^{-u}.
\end{equation}
The velocity of the rear end at this instant can then be written as
\begin{equation}
w = \tanh\left(\frac{ue^u}{l_0} \tau\right),
\end{equation}
meaning that the average proper acceleration is
\begin{equation}
a = \frac{ue^u}{l_0}.
\end{equation}
The force acting on the rod is
\begin{equation}
p = \frac{\rho_0 v}{1 - v} = \rho_0 \, e^u \sinh(u) = \rho_0 l_0 a \, \frac{\sinh(u)}{u} = m_0 a \, \frac{\sinh(u)}{u},
\end{equation}
where $m_0 = \rho_0 l_0$ is the rest mass of the relaxed rod, meaning that for $u \ll 1$ we recover the familiar equation
\begin{equation}
p = m_0 a.
\end{equation}
%
%
%%%%%%%%%%%%%%%%%%%%%%%%%%%%%%%%%%% Section 5 %%%%%%%%%%%%%%%%%%%%%%%%%%%%%%%%%%
%
\section{Bell's spaceships paradox}\label{section5}
\begin{figure}[h!]
\begin{center}
\psfrag{t}{$t$}
\psfrag{x}{$x$}
\psfrag{I}{$I$}
\psfrag{II}{$II$}
\epsfxsize=.5\textwidth
\leavevmode
\epsfbox{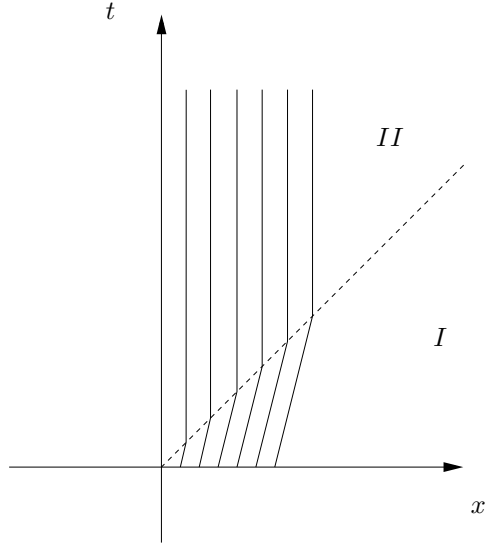}
\end{center}
\caption{Spacetime diagram for a semi-infinite rigid string being held at one end.}\label{pull}
\end{figure}

We start by considering a semi-infinite string in Minkowski spacetime, initially relaxed and moving with velocity $v$ along the $x$-axis, before being held by an irresistible force applied at its end. This situation is dual to that of a semi-infinite rod colliding with an unmovable wall (Section~\ref{section1}), and a similar analysis leads to the initial-boundary value problem
\begin{equation}
\begin{cases}
\Box \lambda = 0 \quad & (t>0,x>0)\\
\lambda(0,x) = \gamma x \quad & (x>0)\\
\dot{\lambda}(0,x) = - v\gamma \quad & (x>0)\\
\lambda(t,0) = 0 \quad & (t>0)
\end{cases}
\end{equation}
whose solution is
\begin{equation}
\lambda(t,x) = 
\begin{cases}
\gamma (x - vt) \qquad & (t>0,x>0,t < x) \\
\gamma (1-v) x  \qquad & (t>0,x>0,t > x)
\end{cases}
\end{equation}
The worldlines of the particles in the string (level sets of $\lambda$) are depicted in Figure~\ref{pull}. We see that there is a rarefaction wave propagating from the holding event $(t,x)=(0,0)$ through the string at the speed of light, represented by the line $t=x$. Before being reached by the rarefaction wave (region $I$ in Figure~\ref{pull}), the points in the string are moving with velocity $v$ and are not stretched:
\begin{equation}
n^2 = | d\lambda |^2 = - v^2 \gamma^2 + \gamma^2 = 1.
\end{equation}

\noindent After being reached by the rarefaction wave (region $II$ in Figure~\ref{pull}), the points of the string are at rest and stretched:
\begin{equation}
n^2 = | d\lambda |^2 = \gamma^2 (1 - v)^2 = \frac{(1 - v)^2}{1 - v^2} = \frac{1 - v}{1 + v} < 1,
\end{equation}
and so
\begin{equation}
p = \frac{\rho_0}2 (n^2 - 1) = - \frac{\rho_0 v}{1 + v} < 0.
\end{equation}

\begin{figure}[h!]
\begin{center}
\psfrag{t}{$t$}
\psfrag{x}{$x$}
\psfrag{I}{$I$}
\psfrag{II}{$II$}
\epsfxsize=.5\textwidth
\leavevmode
\epsfbox{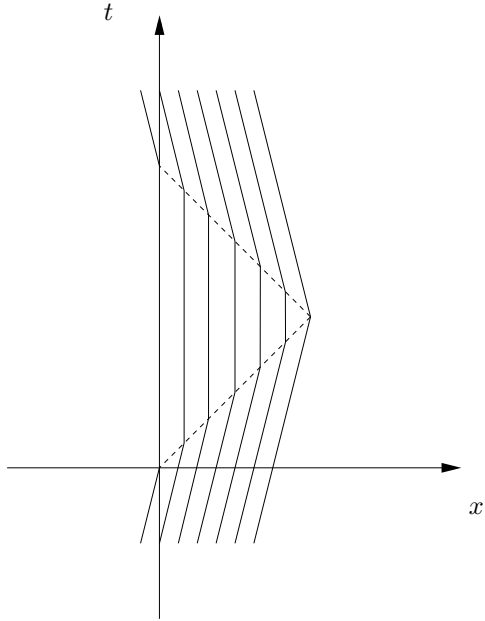}
\end{center}
\caption{Spacetime diagram for a finite rigid string being held at one end.}\label{pull2}
\end{figure}

This solution can easily be adapted to model a finite string being held at one end by an irresistible force. The spacetime diagram for such a situation is depicted in Figure~\ref{pull2}, where we assume that the string was held at the event $(t,x)=(0,0)$. By symmetry, it is clear that as the rarefaction wave reaches the free end of the string a compression wave starts propagating backwards, and the points of the string start moving towards $x=0$ with velocity $-v$. If the string has rest length $l_0 = \gamma l$ then its length at the instant $t=\frac{l}{1-v}$, when it is fully stretched, is easily computed to be
\begin{equation}
l_s = \frac{l}{1-v} = l_0 \sqrt{\frac{1+v}{1-v}} \equiv \frac{l_0}{n}.
\end{equation}

\begin{figure}[h!]
\begin{center}
\psfrag{t}{$t$}
\psfrag{x}{$x$}
\psfrag{I}{$I$}
\psfrag{II}{$II$}
\epsfxsize=.5\textwidth
\leavevmode
\epsfbox{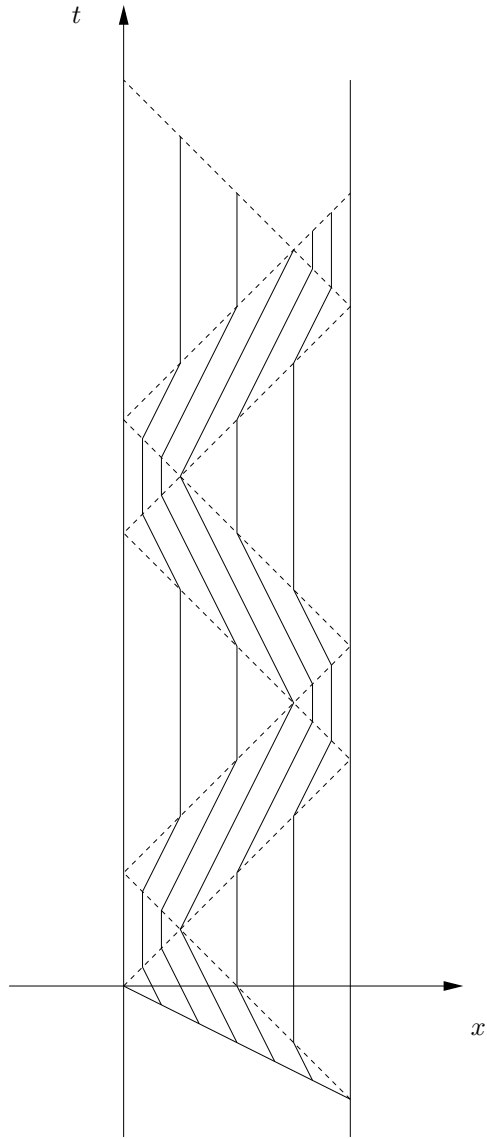}
\end{center}
\caption{Spacetime diagram for a finite rigid string being held at its two ends.}\label{Bell1}
\end{figure}

The solutions describing a finite rod colliding with an unmovable wall or a finite string being held at one end by an irresistible force can be patched together to construct a solution describing a finite string being held at its two ends, as depicted in Figure~\ref{Bell1}. The solution is periodic, and the string cycles through states where it is moving with velocity $-v$ while relaxed, at rest while half compressed and half stretched\footnote{By this we mean that the point where the string switches from compressed to stretched is the midpoint of the relaxed configuration. This is an obvious consequence of the spacetime diagram's gliding reflection symmetry.}, moving with velocity $v$ while relaxed, and at rest while half stretched and half compressed.

\begin{figure}[h!]
\begin{center}
\psfrag{t}{$t$}
\psfrag{x}{$x$}
\psfrag{I}{$I$}
\psfrag{II}{$II$}
\epsfxsize=.8\textwidth
\leavevmode
\epsfbox{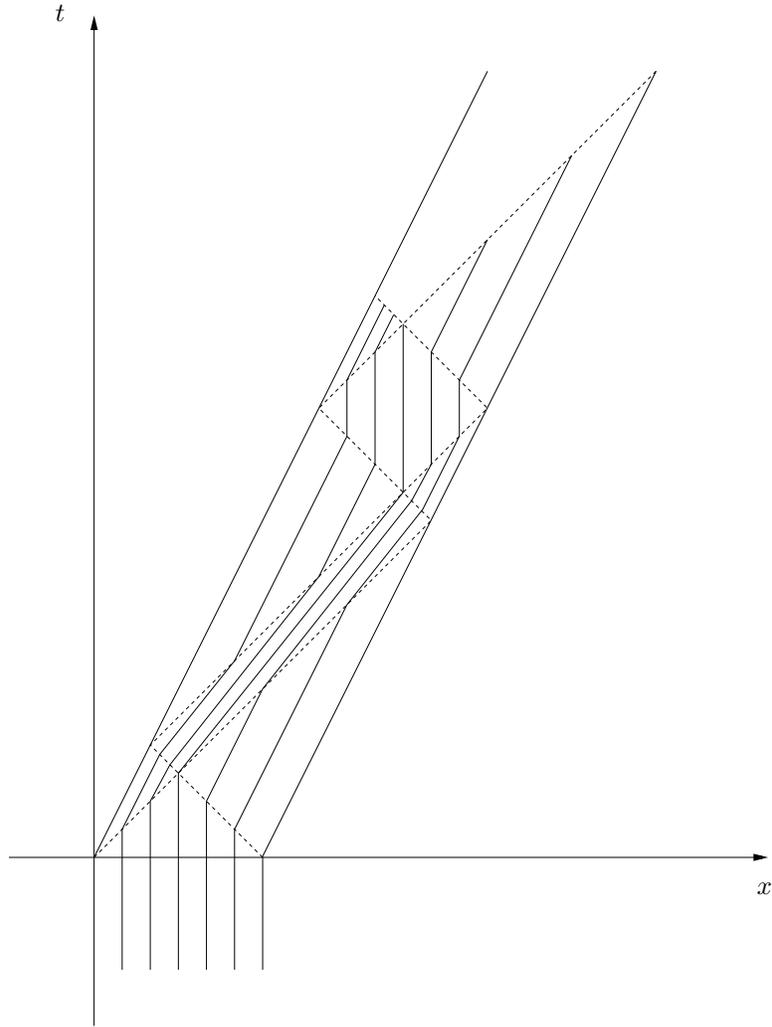}
\end{center}
\caption{Spacetime diagram for Bell's spaceships paradox.}\label{Bell2}
\end{figure}

By changing to the frame which moves with velocity $-v$ with respect to the string's endpoints, we can reinterpret this solution as describing a rigid string initially at rest whose endpoints start moving with velocity $v$ at time $t=0$ (see Figure~\ref{Bell2}). This is a simplified instance of the celebrated Bell paradox \cite{DB59, Bell87}, where two identical spaceships connected by a string start moving simultaneously with the same acceleration profile, thus keeping their separation distance constant. The (apparent) paradox is that the string will stretch, unlike what may be expected at first sight. In the spaceships' initial rest frame, this is because it must compensate the decrease in length due to Lorentz contraction; in the spaceships' final rest frame, this is because the spaceship at the rear is still moving backwards when the spaceship at the front stops moving. Figure~\ref{Bell2}, however, shows that things are not so simple (at least for this model): the string initially becomes stretched at the front and compressed at the rear, until it reaches a state where it is half stretched and half compressed, and is never completely stretched. Moreover, the periodicity of its motion guarantees that it eventually becomes momentarily relaxed and at rest in the spaceships' initial rest frame.
%
%
%%%%%%%%%%%%%%%%%%%%%%%%%%%%%%%%%%% Section 6 %%%%%%%%%%%%%%%%%%%%%%%%%%%%%%%%%%
%
\section{Fishing in a black hole}\label{section6}
As a final example, we analyze the motion of a radial rigid string that has partially crossed the event horizon of a Schwarzschild black hole while still being held from the outside. A related problem was studied in \cite{Brotas04}, but although the calculations seem to be correct the final conclusion is unclear. See also \cite{Egan} for an analysis involving the Rindler horizon and a different elastic law.

\begin{figure}[h!]
\begin{center}
\psfrag{t}{$t$}
\psfrag{x}{$x$}
\psfrag{I}{$I$}
\psfrag{II}{$II$}
\psfrag{x0}{$x_0$}
\psfrag{x-}{$x_-$}
\psfrag{x+}{$x_+$}
\psfrag{xt+}{$\tilde{x}_+$}
\psfrag{r=0}{$r=0$}
\psfrag{r=r0}{$r=r_0$}
\epsfxsize=.7\textwidth
\leavevmode
\epsfbox{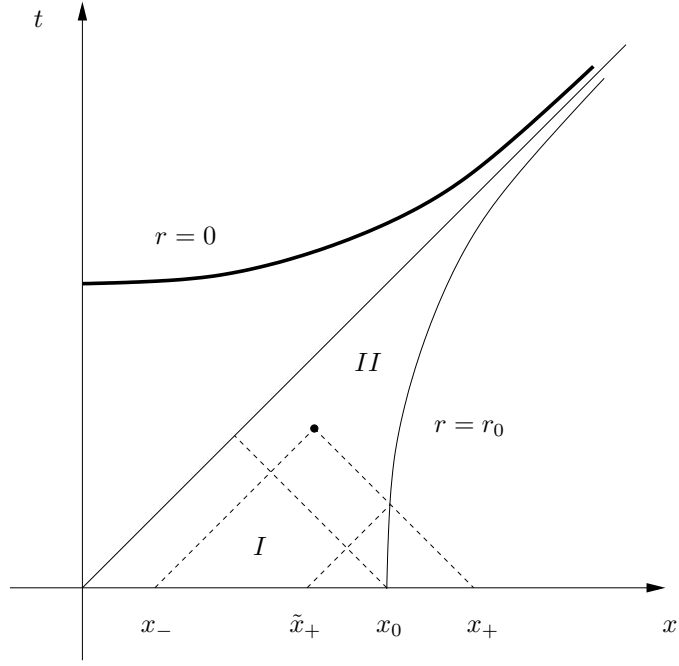}
\end{center}
\caption{Spacetime diagram for fishing in a Schwarzschild black hole.}\label{fishing}
\end{figure}

We consider the totally geodesic $2$-dimensional submanifold of the Schwarzschild spacetime obtained by setting the angular coordinates constant (i.e.~we only study motions along a radial direction). If we choose the Schwarzschild radius to be our length unit, $2M=1$, then the metric of this submanifold in Kruskal-Szekeres coordinates $(t,x)$ is
\begin{equation} \label{Kruskal}
ds^2 = \frac{4e^{-r}}{r} \left( -dt^2 + dx^2 \right),
\end{equation}
where the radius function $r=r(t,x)$ can be computed from 
\begin{equation} \label{r(t,x)}
x^2 - t^2 = (r-1)e^r
\end{equation}
(see for instance \cite{MTW73}). Assuming that the string is being held at $r=r_0$ and that it is initially relaxed and at rest with respect to the stationary observers, we must then solve the following initial-boundary value problem:
\begin{equation} \label{problem_fishing}
\begin{cases}
\Box \lambda = 0 \quad & (t>0,1<r<r_0)\\
\lambda(0,x) = G(x) \quad & (0<x<x_0)\\
\dot{\lambda}(0,x) = 0 \quad & (0<x<x_0)\\
\lambda(x_0 \sinh u, x_0 \cosh u) = G(x_0) \quad & (u>0)
\end{cases}
\end{equation}
Here
\begin{equation}
x_0 = \left( r_0 - 1 \right)^\frac12 e^{\frac{r_0}2}
\end{equation}
is the initial value of the string's endpoint abscissa, and
\begin{equation}
G(y)=\int_0^y \frac{2e^{-\frac{r}2}}{r^\frac12} dx, \qquad \text{ with } \qquad x = \left( r - 1 \right)^\frac12 e^{\frac{r}2},
\end{equation}
is the proper length function along $t=0$ (so that initially $d \lambda^2$ pulls back to the spatial Schwarzschild metric, that is, $n^2 = \left|\grad \lambda\right|^2 = 1$). Although we are solving the wave equation in the region bounded by the event horizon (the line $t=x$) and the timelike curve $r=r_0$ (the hyperbola $x^2 - t^2 = {x_0}^2$), we only require a boundary condition at the latter, because the event horizon is a null line (see Figure~\ref{fishing}).

Using the general form \eqref{wave_soln} of the solution of the wave equation in one spatial dimension\footnote{Note that the Kruskal-Szekeres coordinates $(t,x)$ are conformal, and so the wave equation \eqref{wave_eqn} in these coordinates is just the standard Minkowski wave equation $\ddot{\lambda}=\lambda''$.}, one easily obtains from the initial condition
\begin{equation}
f(x) = g(x) = \frac12 G(x) \qquad (0 < x < x_0),
\end{equation}
and from the boundary condition
\begin{equation}
f(x_0 e^{-u}) + g(x_0 e^u) = G(x_0)  \qquad (u > 0),
\end{equation}
that is,
\begin{equation}
g(x) = G(x_0) - f\left(\frac{{x_0}^2}{x}\right) = G(x_0) - \frac12 G\left(\frac{{x_0}^2}{x}\right) \qquad (x > x_0).
\end{equation}
Therefore the solution to \eqref{problem_fishing} is
\begin{equation}
\lambda(t,x) = 
\begin{cases}
\displaystyle \frac12 G(x-t) + \frac12 G(x+t) \qquad & (0 < x + t < x_0) \\ \\
\displaystyle \frac12 G(x-t) + G(x_0) - \frac12 G\left(\frac{{x_0}^2}{x+t}\right) \qquad & (x_0 < x + t)
\end{cases}
\end{equation}
We see that the behaviour of the string changes according to whether $0 < x + t < x_0$ (region $I$ in Figure~\ref{fishing}) or $x + t > x_0$ (region $II$ in Figure~\ref{fishing}). In region $I$, the string does not yet realize that it is being held at $r=r_0$, and so is in free fall while being stretched by the tidal forces; in region $II$, the information that the endpoint is not free has propagated across the string, and it responds accordingly.

At the horizon we have
\begin{equation}
\lambda(t,t) = 
\begin{cases}
\displaystyle \frac12 G(2t) \qquad & (0 < 2t < x_0) \\ \\
\displaystyle G(x_0) - \frac12 G\left(\frac{{x_0}^2}{2t}\right) \qquad & (x_0 < 2t)
\end{cases}
\end{equation}
In particular, we have
\begin{equation}
\lambda\left(\frac{x_0}2, \frac{x_0}2\right) = \frac12 G(x_0),
\end{equation}
that is, by the time the information that the string is being held reaches the horizon, half the string has already been swallowed by the black hole. It is also clear that
\begin{equation}
\lim_{t \to + \infty} \lambda(t,t) = G(x_0),
\end{equation}
meaning that the whole string will eventually cross the horizon.

It is instructive to compute the stretch factor of the string. For that we set
\begin{align}
& x_- = x - t, \\
& x_+ = x + t, \\
& \displaystyle \tilde{x}_+ = \frac{{x_0}^2}{x_+} \label{tilde{x}_+},
\end{align}
so that the solution to \eqref{problem_fishing} can be written as
\begin{equation}
\lambda(t,x) = 
\begin{cases}
\displaystyle \frac12 G(x_-) + \frac12 G(x_+) \qquad & (0 < x_+ < x_0) \\ \\
\displaystyle \frac12 G(x_-) + G(x_0) - \frac12 G\left(\tilde{x}_+\right) \qquad & (x_0 < x_+)
\end{cases}
\end{equation}
Note that $x_-$ and $x_+$ are the abscissas of the points where the past null cone of $(t,x)$ intersects the $x$-axis. On the other hand, the lines $x+t=x_+$ and $x-t=\tilde{x}_+$ meet on the hyperbola $x^2-t^2=x_+\tilde{x}_+={x_0}^2$, and so $\tilde{x}_+$ can be be found by reflecting the null line $x+t=x_+$ on this hyperbola before intersecting with the $x$-axis (see Figure~\ref{fishing}).

If we use $(x_-,x_+)$ as coordinates, the metric \eqref{Kruskal} is written as
\begin{equation}
ds^2 = \frac{4e^{-r}}{r} dx_- dx_+,
\end{equation}
where
\begin{equation}
x_- x_+ = (r-1)e^r.
\end{equation} 
An easy computation then shows that
\begin{equation}
n^2 = |d\lambda|^2 = r e^r \frac{\partial \lambda}{\partial x_-} \frac{\partial \lambda}{\partial x_+}.
\end{equation}
Therefore in region $I$ we have
\begin{equation}
n^2 = \frac14 r e^r G'(x_-) G'(x_+)
\end{equation}
which we rewrite as
\begin{equation} \label{n^4I}
n^4 = \frac{r^2}{r_- r_+} e^{2r-r_--r_+},
\end{equation}
with $r_\pm = r(0,x_\pm)$. This can be further simplified by using \eqref{r(t,x)} together with
\begin{align}
& x_-^2 = (r_--1) e^{r_-}, \label{r_-(x_-)}\\
& x_+^2 = (r_+-1) e^{r_+}, \label{r_+(x_+)}\\
& x_-x_+ = x^2 - t^2, \label{x_-x_+}
\end{align}
to yield
\begin{equation} \label{main_relation_I}
(r-1)^2 e^{2r} = (r_--1)(r_+-1) e^{r_-+r_+}.
\end{equation}
This allows us to write \eqref{n^4I} as
\begin{equation} \label{n^4VI}
n^4 = \frac{V_- V_+}{V^2},
\end{equation}
where
\begin{equation}
V(r) = 1 - \frac1r
\end{equation}
is the square of the timelike Killing vector field, and $V_\pm = V(r_\pm)$.

As a consistency check, notice that for $t=0$ one has $V_-=V_+=V$ (see Figure~\ref{fishing}), and so $n=1$, that is, the string is relaxed. At the horizon, equation \eqref{n^4VI} cannot be used, since $V=V_-=0$, but the initial expression \eqref{n^4I} with $r=r_-=1$ yields
\begin{equation}
n^4 = \frac{1}{r_+} e^{1-r_+}.
\end{equation}
Thus we see that $n<1$, that is, the string is always stretched at the horizon, and the stretch factor increases with time (as $r_+$ is increasing). 

In region $II$ we have
\begin{equation}
n^2 = - \frac14 r e^r G'(x_-) G'(\tilde{x}_+) \frac{d\tilde{x}_+}{dx_+} = \frac14 r e^r G'(x_-) G'(\tilde{x}_+) \frac{\tilde{x}_+^2}{{x_0}^2},
\end{equation}
which we rewrite as
\begin{equation} \label{n^4II}
n^4 = \frac{\tilde{x}_+^4}{{x_0}^4} \frac{r^2}{r_- \tilde{r}_+} e^{2r-r_--\tilde{r}_+},
\end{equation}
with $\tilde{r}_+ = r(0,\tilde{x}_+)$. Again this can be further simplified by using \eqref{r(t,x)}, \eqref{tilde{x}_+}, \eqref{r_-(x_-)} and \eqref{x_-x_+} together with
\begin{equation} \label{tilde{r}_+(tilde{x}_+)}
\tilde{x}_+^2 = (\tilde{r}_+-1) e^{\tilde{r}_+}
\end{equation}
to yield
\begin{equation} \label{main_relation_II}
\frac{\tilde{x}_+^4}{{x_0}^4} (r-1)^2 e^{2r} =  (r_--1)(\tilde{r}_+-1)e^{r_-+\tilde{r}_+}.
\end{equation}
This allows us to write \eqref{n^4II} as
\begin{equation} \label{n^4VII}
n^4 = \frac{V_- \tilde{V}_+}{V^2},
\end{equation}
where $\tilde{V}_+ = V(\tilde{r}_+)$.

As a consistency check, notice that for $x+t=x_0$ one has $V_+=\tilde{V}_+$, and so equations \eqref{n^4VI} and \eqref{n^4VII} agree, that is, $n$ is continuous. At the horizon, equation \eqref{n^4VII} cannot be used, since $V=V_-=0$, but the initial expression \eqref{n^4II} with $r=r_-=1$ yields
\begin{equation} \label{horizonII}
n^4 = \frac{\tilde{x}_+^4}{{x_0}^4} \frac{1}{\tilde{r}_+} e^{1-\tilde{r}_+}.
\end{equation}
Thus we see that $n<1$, that is, the string is always stretched at the horizon, and moreover $n$ tends to zero along the horizon (as $\tilde{r}_+ \to 1$ and $\tilde{x}_+ \to 0$). At $r=r_0$ one has $\tilde{V}_+=V_-$ (see Figure~\ref{fishing}), and so 
\begin{equation}
n^4 = \frac{V_-^2}{V^2},
\end{equation}
that is, the string is stretched, and the stretch factor increases with time, approaching $+\infty$ (as $V_-$ is decreasing and tends to $0$). 

It is possible to prove that, except for $t=0$, the string is always stretched. For that we change to the coordinates $(r_-,r_+)$; taking the partial derivative of \eqref{main_relation_I} with respect to $r_-$ yields, after some algebra,
\begin{equation}
\frac{\partial r}{\partial r_-} = \frac12 \frac{V_-}{V}.
\end{equation}
It is then easy to compute that
\begin{equation}
\frac{\partial}{\partial r_-} \left(\frac{V_-}{V^2}\right) = \frac1{V^2} \left( \frac1{r_-^2} - \frac1{r^2}\right) > 0
\end{equation}
($r > r_-$ as $r$ increases along outgoing light rays). Since neither $V_+$ nor $\tilde{V}_+$ depend on $r_-$, we see from \eqref{n^4VI} that both in region $I$ and in region $II$ we have
\begin{equation}
\frac{\partial n}{\partial r_-} > 0.
\end{equation}
The level sets of $r_+$ are incoming light rays; since $r_-$ decreases along these rays, we see that $n$ also decreases along them. Because $n=1$ in the initial state $t=0$ and $n<1$ at $r=r_0$, we conclude that $n<1$ throughout (except for $t=0$). 

Similar computations yield
\begin{equation}
\frac{\partial r}{\partial r_+} = \frac12 \frac{V_+}{V},
\end{equation}
and so
\begin{equation}
\frac{\partial}{\partial r_+} \left(\frac{V_+}{V^2}\right) = \frac1{V^2} \left( \frac1{r_+^2} - \frac1{r^2}\right) < 0
\end{equation}
($r < r_+$ as $r$ decreases along ingoing light rays). Since $V_-$ does not depend on $r_+$, we see from \eqref{n^4VI} that in region $I$ we have
\begin{equation}
\frac{\partial n}{\partial r_+} < 0.
\end{equation}
The level sets of $r_-$ are outgoing light rays; since $r_+$ increases along these rays, we see that $n$ decreases along them. So $n$ decreases along any future-pointing causal direction in region $I$.

In region $II$ we can use $(r_-, \tilde{r}_+)$ as local coordinates. By differentiating \eqref{tilde{r}_+(tilde{x}_+)} with respect to $\tilde{r}_+$ we obtain
\begin{equation}
\frac{\partial}{\partial \tilde{r}_+} \left( \tilde{x}_+^2 \right) = \tilde{r}_+ e^{\tilde{r}_+}.
\end{equation}
Using this formula to take the partial derivative of \eqref{main_relation_II} with respect to $\tilde{r}_+$ yields, after some algebra,
\begin{equation}
\frac{\partial r}{\partial \tilde{r}_+} = -\frac12 \frac{\tilde{V}_+}{V}.
\end{equation}
It is then easy to compute that
\begin{equation}
\frac{\partial}{\partial \tilde{r}_+} \left(\frac{\tilde{V}_+}{V^2}\right) = \frac1{V^2} \left( \frac1{\tilde{r}_+^2} + \frac1{r^2}\right) > 0.
\end{equation}
Since $V_-$ does not depend on $\tilde{r}_+$, we see from \eqref{n^4VII} that in region $II$ we have
\begin{equation}
\frac{\partial n}{\partial \tilde{r}_+} > 0.
\end{equation}
The level sets of $r_-$ are outgoing light rays; since $\tilde{r}_+$ decreases along these rays, we see that $n$ also decreases along them. So $n$ also decreases along any future-pointing causal direction in region $II$.

To summarize, we have proved that $n$ decreases along any future-pointing null line (either ingoing or outgoing), and consequently along any future-pointing causal direction. Since $n$ is decreasing and tends to zero along $r=r_0$ (to the future), we conclude that $n$ decreases towards zero (and so the string's tension increases towards infinity) along any future-pointing causal curve that does not cross the event horizon\footnote{This is true also for the event horizon, where $n$ is positive -- see equation \eqref{horizonII}.} nor $r=r_0$. Note that although our mathematical model does not contemplate the string breaking, any physical string will certainly do so.  
%
%
%%%%%%%%%%%%%%%%%%%%%%%%%%%%%%%%%%% Section 0 %%%%%%%%%%%%%%%%%%%%%%%%%%%%%%%%%%
%
\section*{Conclusion}
We conclude by briefly recapitulating our results. We showed that the equation of motion for a rigid rod or string in a two-dimensional spacetime is simply the wave equation. We then solved this equation to analyze the collision of a rigid rod with an unmovable wall, and found that the points in the rod are instantaneously stopped by a compression wave propagating at the speed of light from the collision event. As the compression wave reaches the rear end of the rod, a rarefaction wave starts propagating back towards the front, also at the speed of light, setting the points in the rod in motion away from the wall with minus their initial velocity. This sheds light on the celebrated car and garage paradox: when the rear of the car enters the garage the front is about to hit the back wall in the garage's frame, but it has already hit the wall (and is compressed) in the car's initial frame. The same solution can be used to analyze the motion of a rigid rod being pushed by a constant force. In this case the velocity of the points in the rod is piecewise constant and jumps discontinuously (by small increments under ordinary conditions) as compression and rarefaction waves reflect back and forth along the rod at the speed of light. An even more complicated pattern of compression and rarefaction waves occurs when the endpoints of a rigid string are simultaneously set in motion (which can be seen as a particular case of Bell's spaceships paradox). In this case, unlike what might be expected, the string is never completely stretched, and in fact it even becomes momentarily relaxed. Finally, we analyzed a radial rigid string that had partially crossed the event horizon of a Schwarzschild black hole while still being held from the outside. We found that (i) eventually the whole string will cross the horizon, (ii) the force necessary to hold the string increases indefinitely, and (iii) the tension of the string increases along any future-pointing causal direction.
%
%
%%%%%%%%%%%%%%%%%%%%%%%%%%%%%%%%%%% Section 0 %%%%%%%%%%%%%%%%%%%%%%%%%%%%%%%%%%
%
\section*{Acknowledgments}
I thank Prof.~A.~Brotas for kindly providing a copy of his 1969 PhD thesis on relativistic thermodynamics and continuum mechanics (under L.~de Broglie), as well as some of his early papers. I also thank C.~Botelho for useful discussions and comments. This work was partially funded by FCT/Portugal through project PEst-OE/EEI/LA0009/2013.
%
%
%%%%%%%%%%%%%%%%%%%%%%%%%%%%%%%%%%%%% Bibliography %%%%%%%%%%%%%%%%%%%%%%%%%%%%%%%%%%
%

\end{document}